# Topological frustration of artificial spin ice

Jasper Drisko[1], Thomas Marsh[2] & John Cumings[2]

Frustrated systems, typically characterized by competing interactions that cannot all be simultaneously satisfied, display rich behaviours not found elsewhere in nature. Artificial spin ice takes a materials-by-design approach to studying frustration, where lithographically patterned bar magnets mimic the frustrated interactions in real materials but are also amenable to direct characterization. Here, we introduce controlled topological defects into square artificial spin ice lattices in the form of lattice edge dislocations and directly observe the resulting spin configurations. We find the presence of a topological defect produces extended frustration within the system caused by a domain wall with indeterminate configuration. Away from the dislocation, the magnets are locally unfrustrated, but frustration of the lattice persists due to its topology. Our results demonstrate the non-trivial nature of topological defects in a new context, with implications for many real systems in which a typical density of dislocations could fully frustrate a canonically unfrustrated system.

[1] Department of Physics, University of Maryland, College Park, Maryland 20742, USA. [2] Department of Materials Science and Engineering, University of Maryland, College Park, Maryland 20742, USA. Correspondence and requests for materials should be addressed to J.C. (email: cumings@umd.edu).

Dislocations are topological defects[1] ubiquitous in crystalline materials that can cause a diverse range of phenomena across vastly different systems. Examples of these phenomena include: theoretically predicted ferromagnetic dislocations in an antiferromagnetic lattice[2,3], which have recently been experimentally observed in antiferromagnetic NiO[4]; one-dimensional fermionic excitations in topological insulators[5]; plasticity in metallic alloys[6]; and a recent report on confined structural states at dislocations in an FeMn alloy[7]. The topological nature of dislocations means their presence can be measured far away from the actual defect site by following a closed loop around the dislocation core, resulting in displacement by a lattice constant and requiring an extra vector to complete the loop, known at the Burgers vector. While non-topological defects such as vacancies or substitutions can impart useful and favourable properties for crystals in, for example, semi-conductor engineering, the presence of such a defect is generally unknown in parts of the material far removed from it. However, the non-trivial presence of even a single dislocation could produce long-range topological effects that permeate the crystal. Typical dislocation densities in metals are on the order of $10^5$–$10^{12}$ $cm^{-2}$, while in ceramics they can be much lower, as low as $10^4$–$10^6$ $cm^{-2}$ (ref. 8). For this reason, dislocations generally receive less attention in ceramic materials.

A class of ceramics that has attracted considerable recent scientific interest are the spin ices, which have been shown to exhibit geometric frustration[9,10] in the magnetic moments of rare earth atoms on the pyrochlore lattice of corner sharing tetrahedra, analogous to Pauling's description of hydrogen disorder in hexagonal water ice at low temperatures[11]. A finite residual entropy at low temperature, resulting from a macroscopically degenerate ground state, is characteristic of these fully frustrated systems. Despite measurements confirming a residual entropy in spin ice[9] and implying a disordered and degenerate low-temperature state, a unique ground state has nevertheless been predicted, based on long-range dipolar interactions in the system[12]. However, this ordered state has proven difficult to observe experimentally[9,13] and may be complicated by a competing ground state, based on exchange interactions[14]. Notably, one recent study did observe a partial recovery of entropy, below the Pauling value, for single crystal samples of $Dy_2Ti_2O_7$ thermally equilibrated for very long-time scales[15]. These considerations are of course influenced by the crystalline perfection of the materials, the nature and density of defects, and the precise configuration of spins around these defects—details that are challenging to experimentally determine or control. Artificial spin ices (ASIs) present a possible pathway to resolving these problems through lithographically patterned two dimensional (2D) arrays of nanoscale bar magnets[16–23], in which the precise structure of the lattice may be controlled by design and the interactions are well-described by a dipolar model[23–26]. The ordering principles of ASI have been studied through demagnetizing protocols or, more recently, through thermal activation and annealing[17,18,20–22]. Thermal activation has been successful in observing theoretically predicted long-range ordered states in both the square and kagome ASI geometries[17,21,23], and it is a promising direction for future studies.

An important consideration in the relationship between structure and frustration that has received relatively little attention is the presence of topological structural defects. Topological influences on frustration have been studied in protein folding[27], in the carbon bonding of graphene nanoflakes[28], in nematic liquid crystals[29], and in spin configurations at the (001) surface of antiferromagnetic Cr[30,31]. The work in refs 30,31 is the most relevant to our results as it is related to topological effects in ordered magnetic systems. However, the level of control and lack of direct characterization in these studies leave numerous unanswered questions.

Here, we present a controlled and direct study of the topological frustration, specifically due to topological defects, in otherwise ideal crystals, where our use of the term topological frustration follows that of Araki *et al.*[29]. We introduce controlled topological defects in the form of edge dislocations into thermally active square ASI systems and directly observe the resulting spin configurations on annealing. We observe that the system creates large regions of ground-state antiferromagnetic order, as expected for the square ice geometry. However, the dislocations create domain walls that break this order, and their locations and configurations are not constrained, giving a source of entropy among the observed results. The individual plaquettes of the system are locally unfrustrated, but nevertheless the whole structure remains frustrated due to its topology, leading to a high number of low lying energetic states. Schematics of topological frustration as well as other, more commonly studied types of frustration are shown in Fig. 1a–c.

## Results

**One-dislocation crystals**. Square ASI nanomagnet arrays were originally created to model the six-fold vertex degeneracy in pyrochlore spin ice[9,16]. The system has proven an excellent way to study frustration and its effects, but falls slightly short of a completely analogous model to the pyrochlores due to the inequivalent interaction strengths between the four magnets at each vertex. The variation in the distance between adjacent and opposite spins at each vertex lifts the six-fold degeneracy of the ice-rule state and leads to a ground state given by a perfect tiling of type I vertices (Fig. 1g), which is unique up to a global reversal of spins. We have previously demonstrated perfect ground-state ordering in square ASI samples fabricated from thin films of $FePd_3$ (ref. 21). The samples show ground-state order after an annealing procedure where they are heated above the Curie temperature ($T_C$) of the $FePd_3$ ($\sim 150\,°C$) and cooled slowly to room temperature at a rate of $1\,°C\,min^{-1}$. Our observation of large domains of ground-state ordering in the square geometry is robust and highly reproducible, demonstrating the benefits of $FePd_3$ as a material system for ASI studies. In this work, we have modified the perfect square geometry by introducing edge dislocations into square ASI (see 'Methods' section). Figure 1e presents a transmission electron microscopy (TEM) image of a dislocation in our samples. A Burgers circuit (Fig. 1d) is drawn schematically around the defect site to illustrate its topological nature. Samples are heated to 165 °C in an Ar atmosphere and cooled at a rate of $1\,°C\,min^{-1}$, similar to our previous studies[21]. We image our samples using Lorentz TEM[19] (Fig. 1f) and utilize digital image processing to analyse the resulting images (Fig. 1h,i). We fabricate samples with 36 distinct dislocation geometries, 6 geometries with one dislocation and 30 with two dislocations, and pattern up to 16 crystals of each geometry onto each substrate (for full fabrication details, see 'Methods' section).

Our main results are summarized in Fig. 1i. We find that the dislocation samples show large domains of ground-state ordering, but always have a chain of type II and type III higher energy vertices originating from the dislocation point. These chains are required because of the topological nature of the dislocation. A closed path around the defect is topologically altered, making it impossible to support continuous ground-state order, and thus a frustrated chain must be present. The chains in the single-dislocation geometries always continue to an edge of the finite crystal, and a representative example is shown in Fig. 2, displaying the extended frustration of the system. Other domains and domain walls may be present in the sample as are seen in studies of canonical square ASI[17], but one of these chains must always begin at the dislocation point. Other than this, the frustrated chains and conventional domain walls are locally indistinguishable, exhibiting the same basic phenomena. The lowest energy configuration for either would be the shortest possible, straight chain of only type II vertices, extending in a $<11>$ direction, but this is not generally observed. Type III vertices are almost always present in both the frustrated chains and the conventional domain walls likely because the type III vertices play an essential role in domain wall motion in the square ice lattice[22]. In addition, since a frustrated wall's presence is required by topology, there is only a fractionally small change in its total energy to include type III vertices and take a longer, meandering path through the lattice. It appears that both such walls are at least partly entropically driven, as opposed to purely energetically driven, as the lowest energy configurations do not dominate. We note that a vacancy, a single missing magnet without lattice distortion, is not a topological defect and does not induce frustration or nucleate a domain wall; a square ASI lattice can still display perfect order with a single spin removed (Supplementary Fig. 1). In addition, we observe that repeated annealing of the same crystal results in different domain configurations (Supplementary Fig. 2). Finally, topological frustration is distinct from vertex frustration, which is found, for example, in the shakti lattice[20] because vertex frustration is only due to the geometry of the system, not its topology.

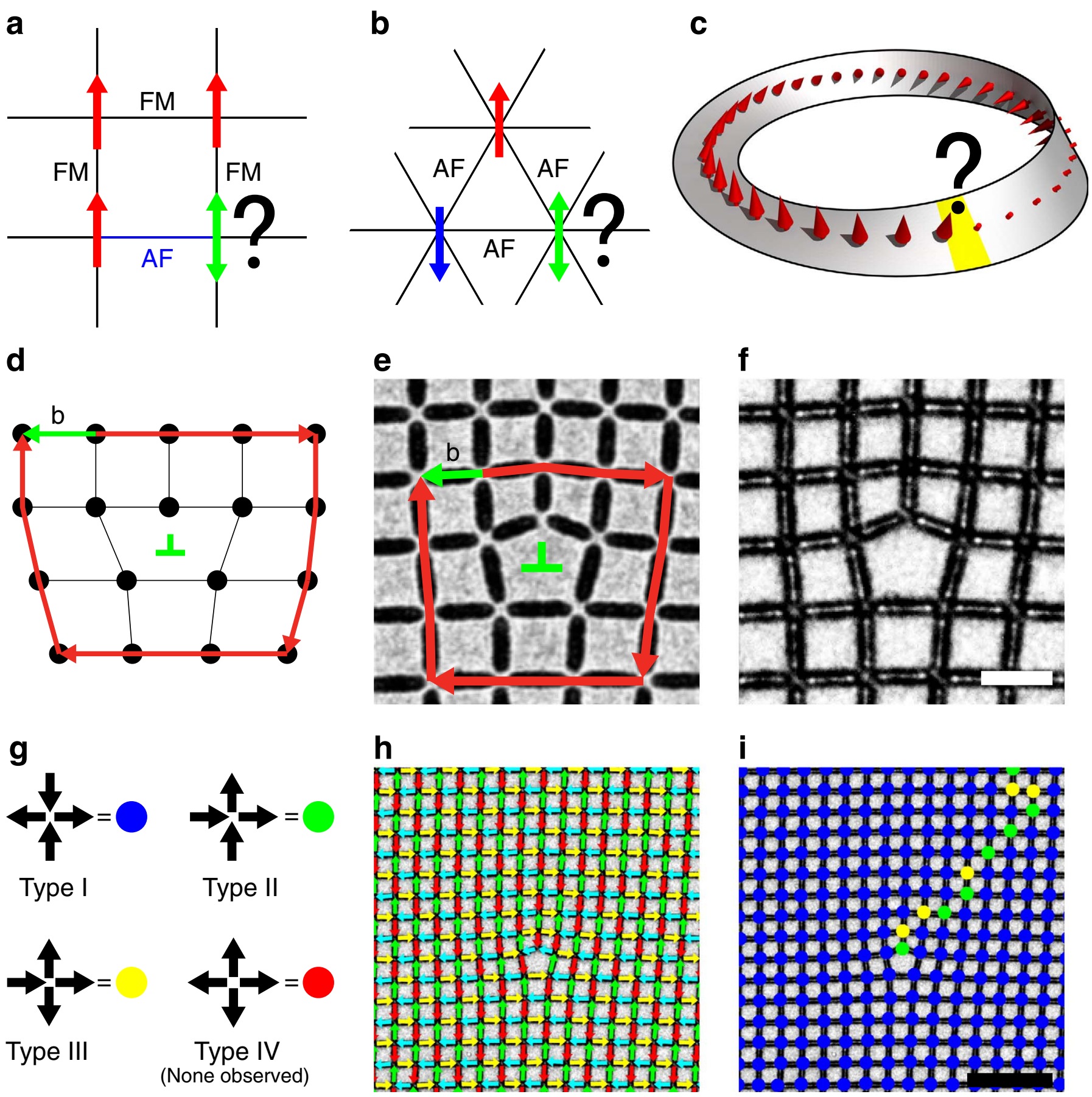

**Figure 1 | Topological frustration.** (**a**–**c**) Schematics of different origins of frustration. (**a**) Disorder frustration: disordered bonds between Ising spins on a square lattice cannot be satisfied with any spin configuration, leading to frustration. (**b**) Geometric frustration: antiferromagnetically (AF) coupled Ising spins on a triangular lattice are frustrated due to the geometry of the system. (**c**) Topological frustration. Ferromagnetic (FM) spins perpendicular to a Mobius strip are locally unfrustrated, but the system is still frustrated due to its topology. (**d**) Burgers circuit and vector around an edge dislocation. A loop around the dislocation (red arrows) with an equal number of lattice constants in each direction cannot be completed without the addition of the Burgers vector (green arrow). Also shown is the conventional 'T' notation (green) representing the dislocation location and direction. (**e**) In-focus TEM image of a dislocation point defect in square ASI. The Burgers circuit and vector are overlaid in the image. (**f**) Lorentz contrast TEM image of the same defect site as in **e**. The asymmetry in contrast across each magnet indicates the direction of its magnetization. For more details see ref. 9. Scale bar, 500 nm. (**g**) Vertex types in square ASI and the legend for vertex map images in **i** and throughout the manuscript. There are 16 possible configurations for 4 spins meeting at each vertex in the square lattice that are divided into four vertex types I–IV with increasing energy. A characteristic spin configuration is given for each type, with the remaining configurations enumerated by four-fold rotation and global spin flip symmetry operations. Two and three spin vertices are simply mapped onto the corresponding four-spin type, excluding the highest energy, type IV vertex because we do not observe any in our samples. (**h**) Lorentz contrast TEM image of a sample that has been annealed with arrows overlaid indicating the direction of each macro spin. (**i**) Vertex map of the same image in **h** representing the same information, but with the domain wall clearly defined by the type II and type III vertices in a single domain of ground-state order. Scale bar, 2 μm.

**Two-dislocation crystals**. For crystals with two dislocations, the two defects generally have separate domain walls that each meander to the edges of the crystal (Fig. 3a). Occasionally, a single domain wall will connect the two dislocations, as the inclusion of two topological defects actually allows for the possibility of a larger continuous ground-state domain (Fig. 3b). From our observations, we compute the probability of domain walls not connecting dislocations in our samples for 30 different two-dislocation geometries (Supplementary Fig. 3) and plot the probability versus the distance between the two dislocations in Fig. 4. An annotated version of Fig. 4 indicating the geometries that contribute to each data point is found in Supplementary Fig. 3h. We find that for a spacing of $>10$ lattice constants, it is extremely rare for a domain wall to connect the two dislocations. At spacings of 5–10 lattice constants, there is still a low probability for a domain wall to span the defects. At 5 lattice constants apart, there is a sharp decrease down to 1 lattice constant spacing. Two points should be noted about this behaviour. The first is that even when the defects are 1 lattice constant apart, there is still a reasonable rate for the two defects to behave independently and nucleate completely separate domain walls. The second is the transition from 1 to 5 lattice spacings shows the emergence of defect independence and at around a separation of 5 lattice constants, the system enters a regime of nearly independent defects. Overall, we find the presence of a domain wall at a dislocation to be extremely robust. We studied five samples fabricated from different depositions of $FePd_3$ and collected over 1,600 images of annealed crystals, and in every image, there is a domain wall originating from every dislocation and connecting to either the edge of the crystal or to another dislocation.

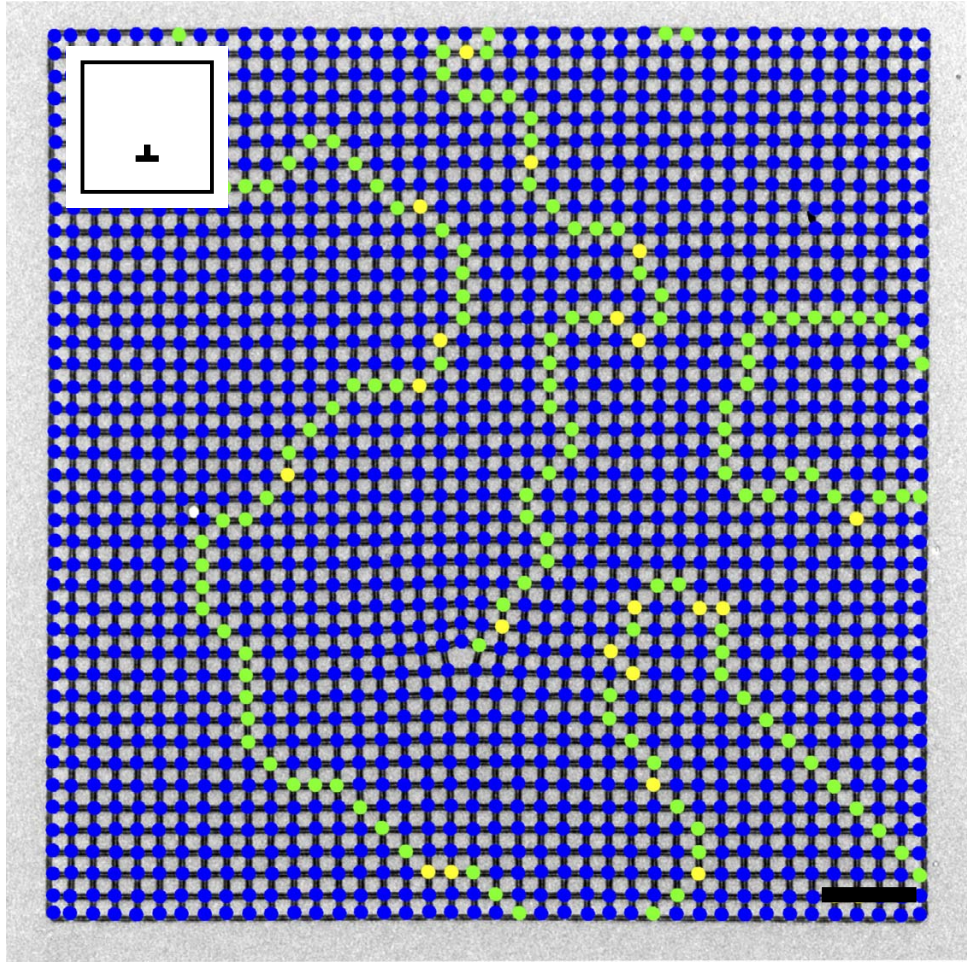

**Figure 2 | Vertex map of a crystal with one topological defect.** Lorentz contrast TEM image with vertex types indicated as in Fig. 1g. A meandering vertex chain originates from the dislocation among a few ground-state domains separated by traditional domain walls. Inset: sample schematic showing location and direction of dislocation. Scale bar, 2 μm.

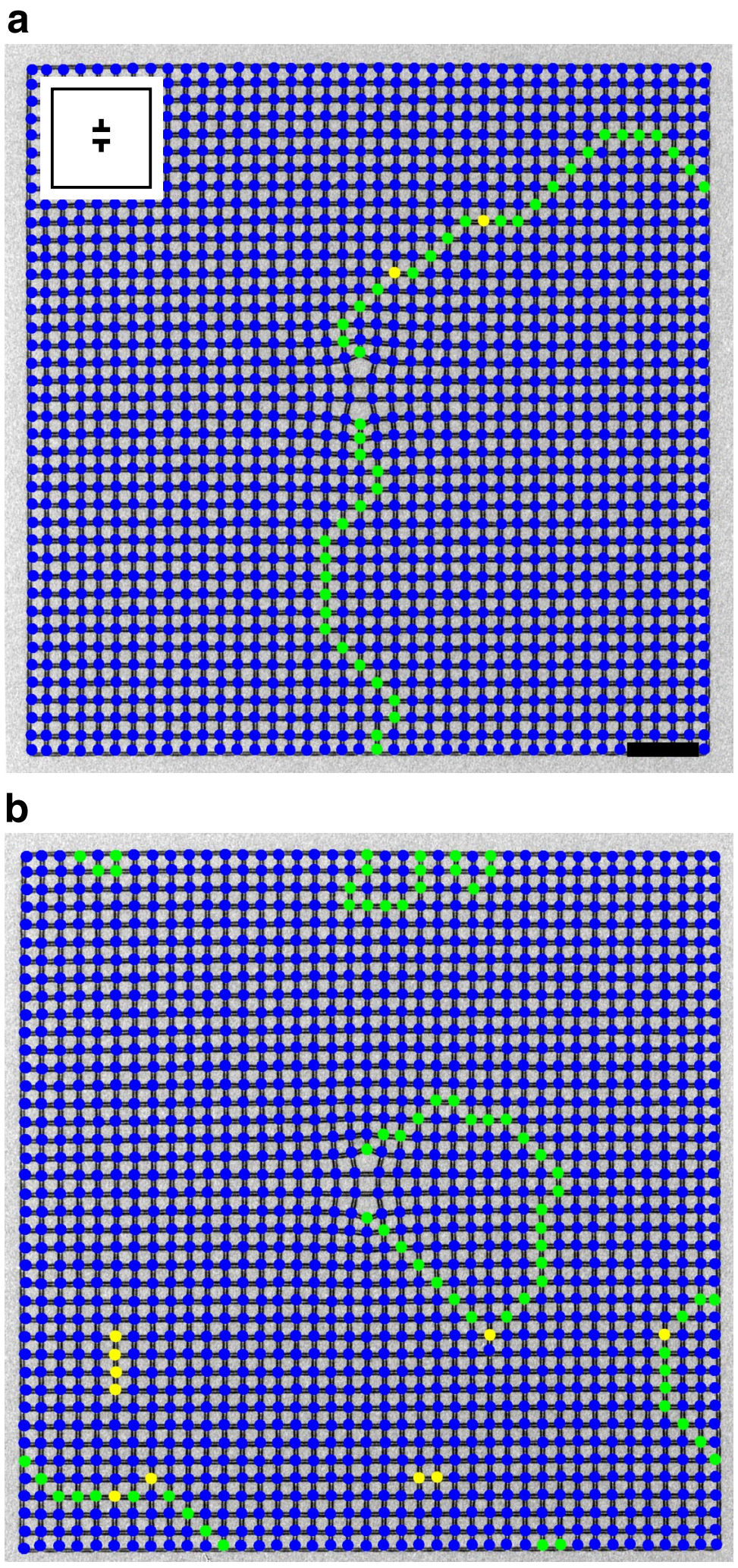

**Figure 3 | Vertex map of crystals with two topological defects.** Lorentz contrast TEM images with vertex types indicated. (**a**) Domain walls begin at each dislocation and end at the edges of the crystal. (**b**) A single domain wall connects the two dislocations. Inset shows a schematic representation of the dislocation locations and direction. Scale bar, 2 μm.

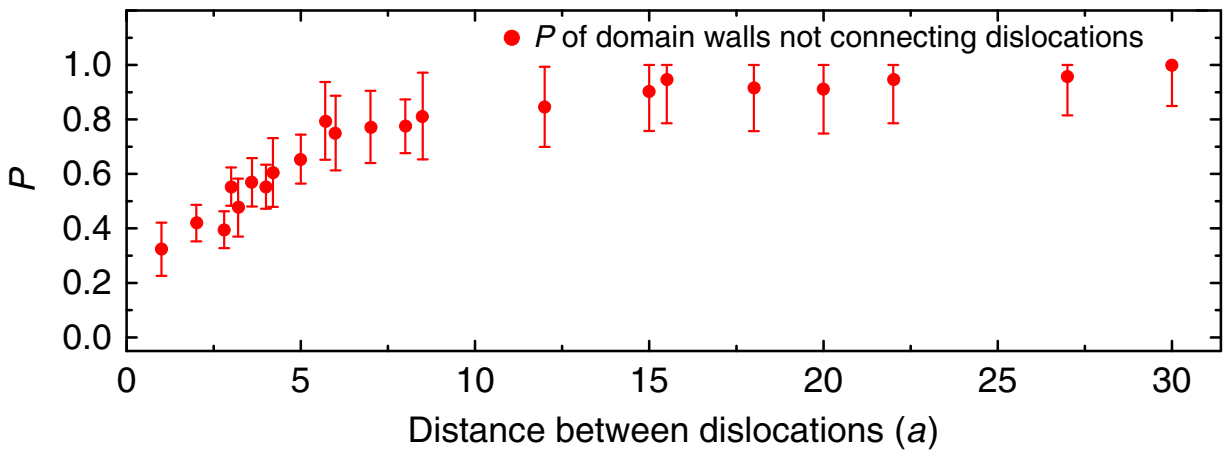

**Figure 4 | Probability of domain walls not connecting dislocations.** Probability of domain walls not connecting dislocations versus distance between dislocations in crystals with two topological defects. Some data points include two or three geometries of equal distance apart. Distances are expressed in units of the lattice constant $a \approx 500$ nm. Error bars are one s.d. and are calculated from counting statistics. Each geometry includes $\sim 40$ crystals.

**Numerical modelling**. We also perform simulations of our system to gain greater understanding of the behaviour. We use a kinetic Monte Carlo technique, which is known to be an accurate method for modelling thermal ASI systems[18,21], and we include a small amount of disorder in the widths of individual magnets in the model (see 'Methods' section). These variations in the samples are due to artifacts from lithography in fabrication and have shown to be important in matching with previous experimental results[21]. We find remarkably similar behaviour between our simulations and TEM data, where the dislocation points always terminate a domain wall, but they are surrounded by large domains of ground-state order. Monte Carlo results also show domain walls separating spontaneous ordinary domains and occasionally connecting two dislocations, consistent with our experimental observations. We also note that some simulations (not shown) with less disorder tend to produce lattices with larger regions of perfect ground-state ordering and fewer canonical domain walls. The simulations shown here include a representative amount of disorder for our experimental crystals, though naturally some experimental crystals will have higher or lower degrees of disorder and will display correspondingly larger or smaller regions of ground-state order surrounding the topologically required domain walls. We show the steady state behaviour of some of our simulations in Fig. 5 and movies of the evolution of the simulations in the Supplementary Movies 1 and 2. In the movies, the lattices start in a randomized state and achieve most of their ordering relatively quickly, first eliminating the highest energy type IV vertices and progressing to mainly type I ground-state order. Once the sample has reached a ground-state domain-dominated phase, the domain walls slowly fluctuate and move around, often shrinking some of the smaller domains. In the domain wall motion, we note fluctuations of type III vertices very similar to previous Monte Carlo simulations of perfect square geometry ASI[22]. We note that topological frustration, while it derives from crystalline disorder, is fundamentally different from standard disorder-induced frustration. Here, the system shows extended frustration as the topology of the defect causes long-range effects in the lattice. In 2D ASI, the dislocation is a 0D point defect, but it promotes itself into a 1D line of frustration. In a 3D material, dislocations are 1D defects, and this would result in an extended 2D surface of frustration of the system.

## Discussion

An important question remains about how our observations relate to dislocations in 3D crystalline materials and the outstanding issue of whether a unique ground state could be realized in spin ice materials. We consider previous measurements of residual entropy in spin ice in the context of sample preparation and dislocation density. The material studied in the seminal report by Ramirez *et al.*[9] was a pressed powder sample, which should result in a high density of dislocations both due to damage during compaction and also due to grain boundaries in the resulting structure. This material, with a presumably high density of dislocations, showed a level of entropy equal to the Pauling ice value, but the $2 \times 1 \times 1\ mm^3$ single crystal studied by Pomaranski *et al.*[15] showed less entropy, a small amount below the Pauling value. Pomaranski *et al.* also studied powder samples, which showed similar spin-freezing and frustrated behaviour as the samples in ref. 9. We postulate that the topological frustration due to a high density of dislocations in both powder samples could be complicating and slowing ground-state ordering, while the single crystal in ref. 15 could be showing a slight emergence of the ground state because it contains far fewer topological defects. However, it is important to note that larger single crystals maintain lattice coherence over greater distances and are thus actually more susceptible to topological frustration. As an example, a cubic crystal of side length $N$ unit cells and volume $N^3$ unit cells could need only on the order of $N_d = N$ dislocations to be fully frustrated, if the emanating frustrated surfaces are non-interacting, similar to the domain walls we observe in our specimens—we discuss this in more detail below. Under this assumption, the dislocation density would be $\rho_d = N_d/N^2$, and the density required to frustrate a system scales as $1/N$, implying that larger single crystals are actually more likely to be impacted by the presence of dislocations than smaller crystals. This is counter to the conventional expectation that larger single crystals show more intrinsic ground-state behaviours. On the contrary, smaller single crystals, or single crystals with lower densities of dislocations, should be better able to reach a fully ordered ground state when topological frustration is present.

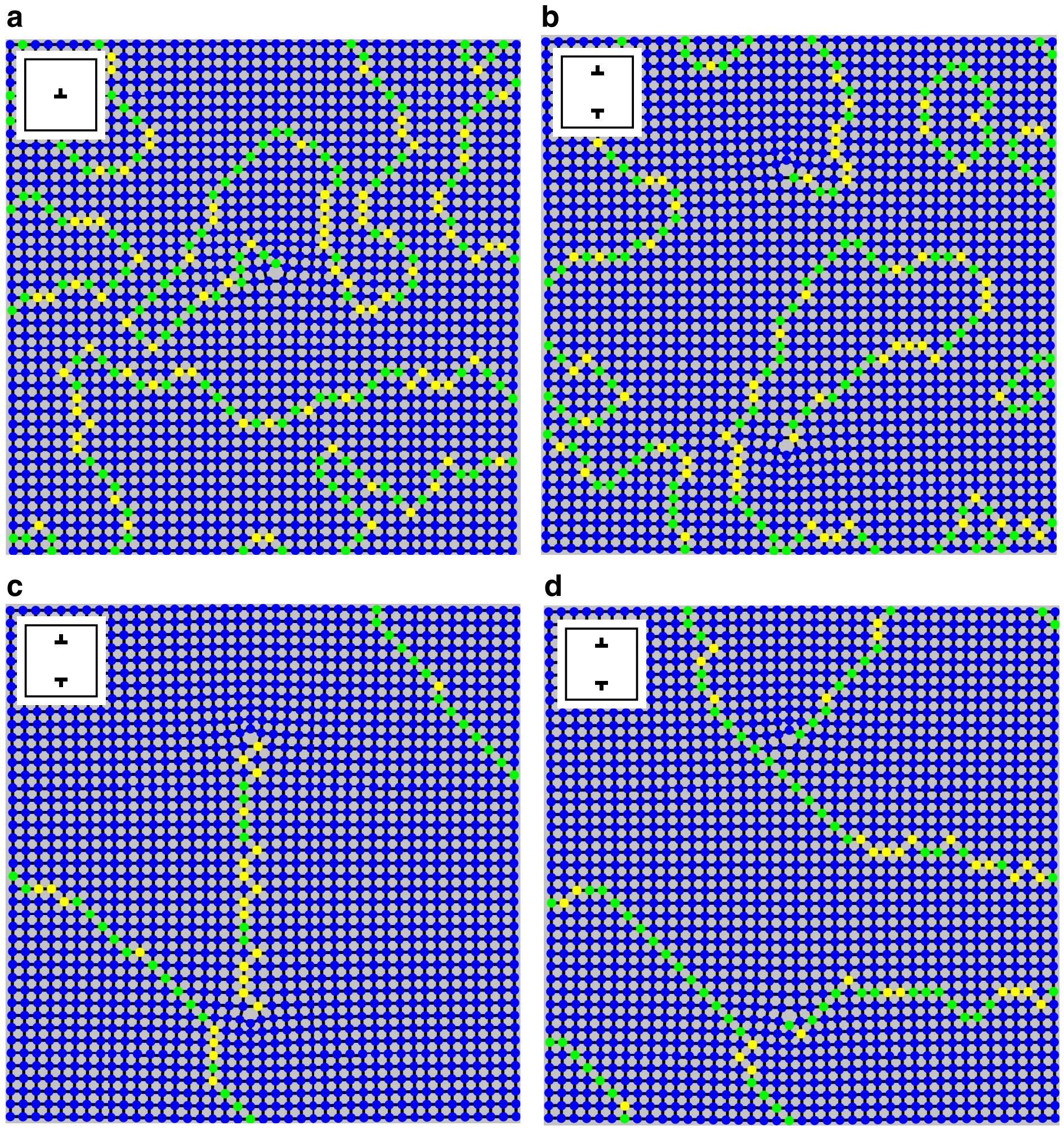

**Figure 5 | Monte Carlo simulations.** Final frames of kinetic Monte Carlo simulations with one (**a**) and two (**b–d**) dislocation lattices. The domain wall patterns observed are remarkably similar to our experimental data with dislocations always nucleating a domain wall that extends to the edge of the finite crystal or to another dislocation among canonical ground-state ordering and domains. The spread in widths used in these runs are 5 Å (**a**), 6 Å (**b**), 6 Å (**c**) and 6 Å (**d**) (see 'Methods' section).

Implications of topological frustration could extend far beyond the systems discussed above, and we first lay out a framework for such considerations as follows. Many materials, simple and exotic alike, display ordered states that can generally be characterized by an order parameter and associated ordering vector, $\mathbf{q} \neq \mathbf{0}$. If the topological Burgers vector of a defect is perpendicular to a given $\mathbf{q}$, or contains a complete wave of the order parameter, with $\mathbf{q} \cdot \mathbf{b} = 2\pi n$, where $n = 0$ or an integer, the system will not be frustrated. However, if these conditions do not hold, the system will exhibit topological frustration. In our square ASI, there are two $\mathbf{q}$ vectors corresponding to the doubly degenerate ground state, $\mathbf{q} = \pm(\pi/a)[11]$, which tile type I vertices onto the lattice. Our choices of Burgers vectors are in the [10] or [01] directions, thus $\mathbf{q} \cdot \mathbf{b} \neq 2\pi n$, and the system will always be topologically frustrated. In the 3D pyrochlore spin ices, there are 6 $\mathbf{q}$ vectors for 6 realizations of the predicted long-range ordered state, $\mathbf{q} = (2\pi/a) < 100 >$[12], as shown in Fig. 6a. To our knowledge, no prior studies have identified the density or nature of dislocations in pyrochlore samples, but it would be straightforward and instructive to analyse their expected influence. The pyrochlore structure is derived from the fluorite structure, which readily forms dislocations with Burgers vectors along one of the primitive FCC lattice vectors, $(a/2) < 110 >$[32], as shown in Fig. 6b. For each possible Burgers direction, 4 out of the 6 dipolar ground-state $\mathbf{q}$ vectors will not be perpendicular to the Burgers vector and will thus be necessarily frustrated, producing a domain wall surface emanating from each such dislocation line, even in the lowest energy state. If a crystal contains dislocations with all FCC primitives, then all 6 ground states will exhibit such effects, and the specimen could be prevented from finding a unique ground state, a possibility we discuss further below.

A recent and thorough analysis of $Dy_2Ti_2O_7$ results speculates that random disorder, in the form of stuffed spins or oxygen vacancies, could be the cause of discrepancies between various specific heat measurements on the material[14]. Stuffed spins have indeed been shown to influence magnetic relaxation times in these systems[33], but the actual disorder and the true source of the specific heat discrepancies remain unresolved. We propose that topological frustration due to dislocations may be an additional contributing source of random disorder, and we consider the implications as follows. As noted above, we have assumed in the prior discussion that the domain walls due to dislocations are non-interacting, as appears to be the case in our studies of 2D ASI. This appearance may be influenced by dynamical arrest, where the domain configurations become locked in our experiments due to local disorder, and the motion of the domain walls slows down significantly. This effect is apparent in

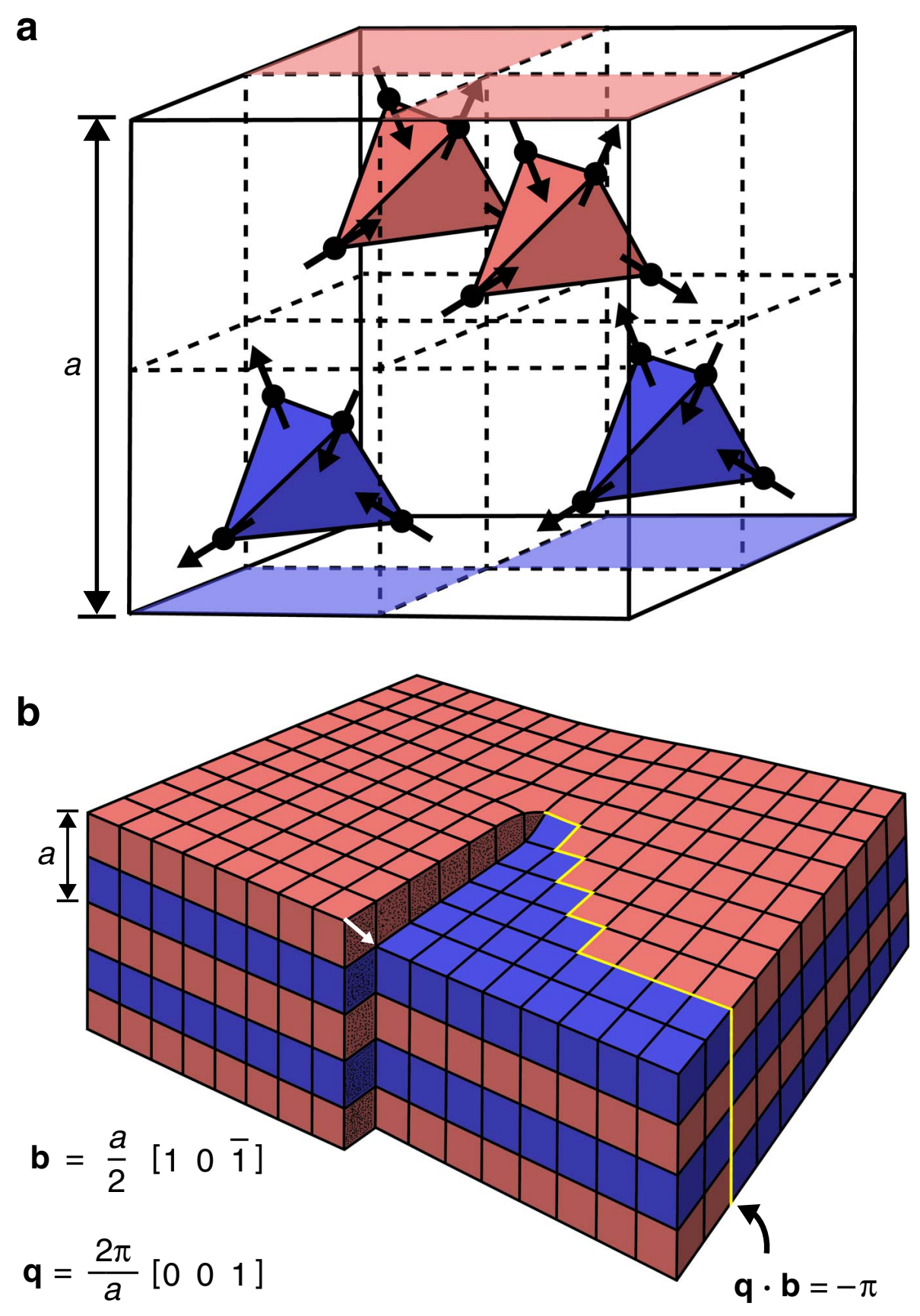

**Figure 6 | Dislocations in pyrochlore spin ice.** (**a**) Reduced unit cell showing the locations of the rare earth atoms and their magnetic moment directions in the possible $\mathbf{q} = (2\pi/a)[001]$ dipolar ground state. This state has alternating planes of ordered tetrahedra, here coloured red and blue, corresponding to the planes of ordering in (**b**). (**b**) Schematic showing how a dislocation in pyrochlore spin ice could disrupt ground-state ordering. Each cube is 1/8 the pyrochlore unit cell shown in **a**. This example has a dislocation with a Burgers vector (shown as a white arrow) in the $(a/2)[10\bar{1}]$ direction, which is a primitive basis vector for the FCC parent structure of the pyrochlore lattice so that this does not cause any structural mismatch away from the dislocation. This dislocation does create a domain wall in the $\mathbf{q} = (2\pi/a)[001]$ ground-state order parameter, highlighted in yellow. The location of this domain wall is arbitrary, but it must be present. Note that $\mathbf{q} \cdot \mathbf{b} \neq 2\pi n$ and thus the system is topologically frustrated.

Supplementary Movies 1 and 2, and related freezing phenomena are also observed in pyrochlore spin ices[34,35]. Despite this, domain walls separating ground-state regions should still carry an additional energy per unit area, producing a surface tension that would tend to minimize the significance of the domain walls, in the limiting case producing a single configuration with only minimal entropy. If the system remains sufficiently dynamic, the mere presence of dislocations may thus not prevent the formation of extended ground-state regions, separated by discrete domain walls. Since the domain walls carry a $\pi$ phase shift of the order parameter (an antiphase boundary), the domain walls may simply find a minimum-energy configuration connecting two nearby dislocations. On the other hand, if the system is otherwise frozen, even a dislocation-free crystal may still show residual entropy without long-range order. Thus, the role of dislocations is not necessarily simple. However, we point out that crystalline dislocations—since they are topological defects—behave differently in three dimensions than in the 2D systems we study here.

In 2D crystals, a dislocation is a point defect, but in three dimensions it is a topological line defect and must either close in a loop or extend to the edges of a crystal. If line dislocations in a 3D crystal occur on a simple parallel lattice, then a trivial arrangement of domain walls may connect them, producing a straightforward ground-state configuration. However, an arbitrary configuration of dislocations could produce complicated arrangements of domain walls, which may not even be unique. In this case, a material which may otherwise be able to find a ground state could see a robust frustration simply due to the presence of topological defects. Furthermore, it has been proposed that a second competing ground state with longer ordering period may play a role in spin ice behaviour[14]. For this order parameter, a dislocation would not produce a simple antiphase boundary but instead a more complicated phase shift. Thus, dislocations may play an even more significant role than described above. The possible influence of topological frustration on spin ordering in these systems is therefore not straightforward and warrants more attention in future experimental and theoretical investigations.

Similar frustration could be present in other systems with $\mathbf{q}\neq\mathbf{0}$ order parameters. Prime examples may include spin-density-wave materials[36], such as antiferromagnets and ferrimagnets, antiferroelectrics, such as $SrTiO_3$ (ref. 37), charge-density-wave materials[38] and pair-density-wave superconductors[39]. In cases where the dislocation densities are low, or where the frustrated domain walls have stronger surface tension between dislocations, the frustration may not prevent the formation of the order parameter, but rather it will only degrade it or delay it somewhat. In such cases, we expect topological frustration may still thermally broaden the phase transition, degrading the formation of the $\mathbf{q}\neq\mathbf{0}$ order parameter. Topological defects may even play a role in the frustration of ice-XI, the thermodynamic ground state of water not known to form under typical experimental conditions[40]. In general, the effects of topological frustration could be observed in a wide variety of materials systems, and ASI could serve as a valuable platform for future studies to develop these general considerations.

## Methods

**Sample fabrication.** To define our lithography patterns, square lattices are first modelled as a network of connecting springs. The springs are all of equal spring constant and attempt to keep the lengths between neighbouring nodes the same. We also include a term that aims to keep the angles between connections coming out of each node the same, for example, 90° for four connections or 120° for three connections. We remove a chain of nodes starting from a given point, doubling the spring lengths along the chain of removed points. We then let the system relax and after it has equilibrated, we record the $x$ and $y$ location of all nodes. We then use these relative coordinates to as a guide for our lithography patterns. A movie of the relaxation process is given in Supplementary Movie 3. A description of all geometries studied is given in Supplementary Fig. 3.

Our ASI arrays are fabricated from 23 nm thick $FePd_3$. We have reported film grown and fabrication details previously[21]. Elements are 120 nm wide and the nominal lattice constant is 500 nm, though this varies due to the relaxation in the connected spring model. Element lengths are given by subtracting a fixed amount from each node, with the three connected vertices treated by subtracting a slightly larger amount to prevent elements from touching.

**Heating experiments and imaging.** After fabrication, samples are heated to 165 °C in an Ar atmosphere and cooled at a rate of 1 °C min$^{-1}$. The resulting magnetic configurations are imaged in a JEOL JEM 2100 $LaB_6$ microscope. A degaussing procedure is run on the microscope's objective lens before inserting the sample to remove any remnant field that could bias the magnets. The field at the sample is measured to be <1 G after this procedure. Lorentz contrast images are then taken and analysed with the aid of computerized automatic image processing.

**Numerical methods.** Monte Carlo simulations are performed using a kinetic Monte Carlo technique. We use the same procedure to define our simulation lattices as we do our lithography patterns. Simulations have the same number of spins, same size spins, and same boundary conditions as experimental samples. In kinetic Monte Carlo simulations, each element is assigned a flip rate given by

$$\tau^{-1} = \nu_0 \exp\left(\frac{E_0 + \Delta E}{k_B T}\right) \quad (1)$$

where $\nu_0$ is a prefactor, $\Delta E$ is the change in energy given by nearest neighbour dipolar coupling, $T$ is the temperature and $E_0$ is the intrinsic energy barrier. We are less concerned with the parameters $\nu_0$ and $T$ as they affect every rate in the same way and are generally used to evolve time in the model, where we are mainly interested in the long-time steady state configuration of the system. After assigning rates to each spin, the rates are summed and used to create a probability distribution so that elements in energetically unfavourable arrangements have a higher probability of flipping. We then randomly pick one element weighted by the probabilities, flip it, and recalculate all the flip rates. As in our previous studies[21], we include a small amount of disorder in the form of width variations in individual elements. We use Gaussian distributions with average 120 nm and characteristic spreads on the order of 1–10 Å. This width disorder factors into the simulation in two ways: first, the intrinsic energy barrier, $E_0$, is approximated by the shape anisotropy energy of the element, which is the product of the demagnetizing factor, volume, and the magnetization of the material squared[18,41]. Both the demagnetizing factor and the volume are affected by the range of element widths. Second, the disorder enters through the magnetic dipolar coupling, given by

$$E = -\frac{\mu_0}{4\pi|\mathbf{r}|^3}\left(3(\mathbf{m}_1\cdot\hat{\mathbf{r}})(\mathbf{m}_2\cdot\hat{\mathbf{r}}) - \mathbf{m}_1\cdot\mathbf{m}_2\right) \quad (2)$$

where $\mathbf{r}$ is the vector connecting the midpoints of the dipolar spins pointing from 1 to 2 and $\hat{\mathbf{r}}$ is the unit vector along $\mathbf{r}$. The magnetic moments $\mathbf{m_1}$ and $\mathbf{m_2}$ are the magnetization of the material times the volume of the element and are thus also affected by the width disorder. The vector orientations of the individual magnets in the relaxed lattices are also taken into account with this dipolar coupling. The nanomagnets' small deviations from regular square lattice orientations slightly broaden the four vertex energy levels into bands, but not enough so that they overlap or reorder. Movies of the simulations are given in the Supplementary Information.

**Data availability.** The data that support the findings of this study are available from the corresponding author upon reasonable request.

## Acknowledgements

This work was supported by NSF CAREER Grant No. DMR-1056974. We also acknowledge the support of the Maryland NanoCenter and its AIMLab and FabLab.

## Author contributions

J.C. and J.D. conceived of the experiment. J.C. supervised the experiments, simulations and data analysis. J.D. deposited thin films, fabricated ASI samples, performed the experiments and ran Monte Carlo simulations. T.M. processed Lorentz TEM images and analysed data with J.D. J.D. wrote the manuscript with input from all authors.

## Additional information

**Supplementary Information** accompanies this paper at http://www.nature.com/naturecommunications

**Competing financial interests:** The authors declare no competing financial interests.

**Reprints and permission** information is available online at http://npg.nature.com/reprintsandpermissions/

**Publisher's note:** Springer Nature remains neutral with regard to jurisdictional claims in published maps and institutional affiliations.

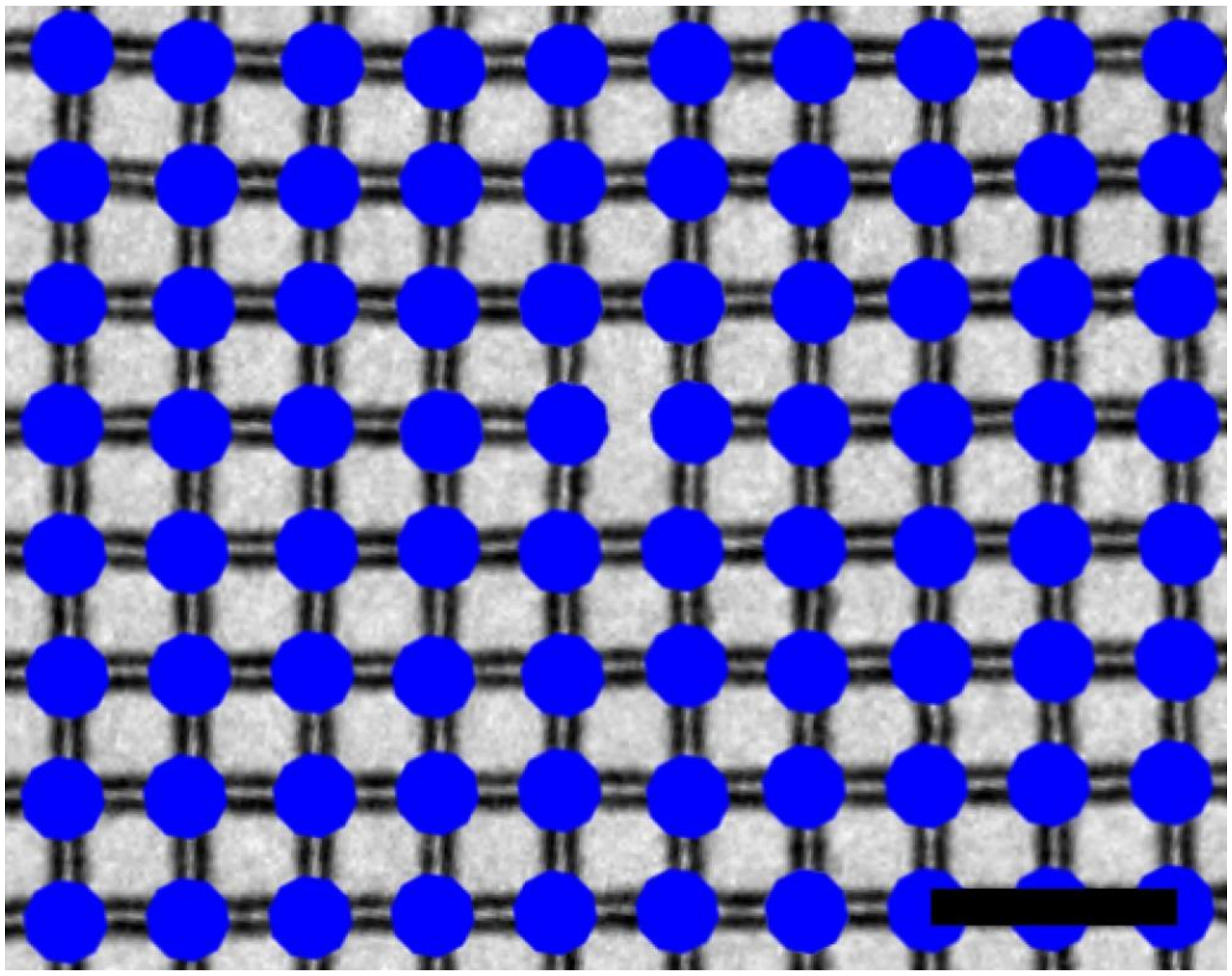

**Supplementary Figure 1 | Vertex map of a non-topological defect.** A simple vacancy, a single missing spin in an otherwise perfect square lattice, does not nucleate a vertex chain or result in topological frustration. Long range ground state ordering is not affected by the presence of a vacancy since it is not a topological defect. Scale bar is 1 μm.

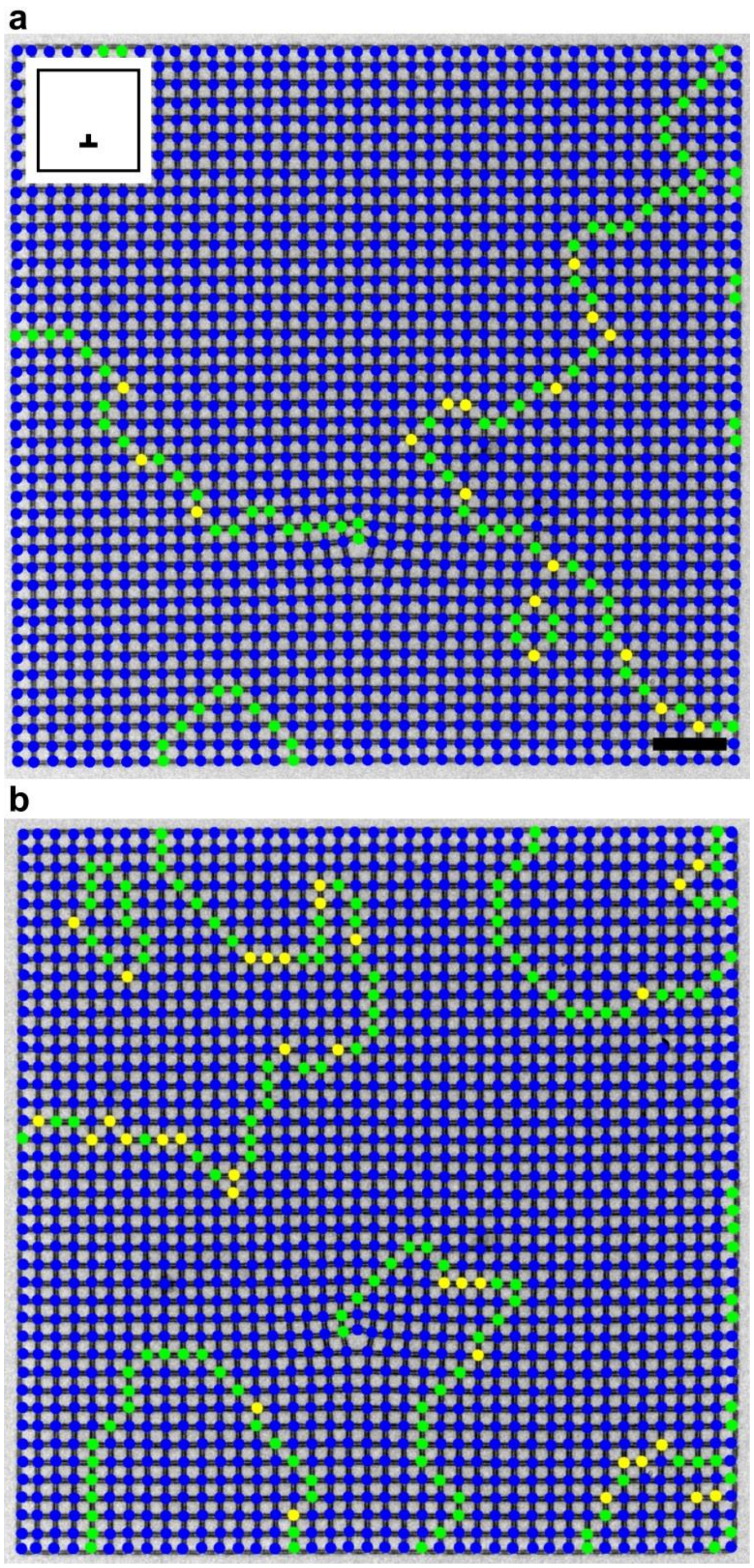

**Supplementary Figure 2 | Repeated annealing of the same crystal.** The same crystal is annealed twice under the same conditions and results in different domain configurations. The quenched disorder in the sample gives qualitatively similar, but not exactly reproducible ordering of vertices. Scale bar is 2 μm.

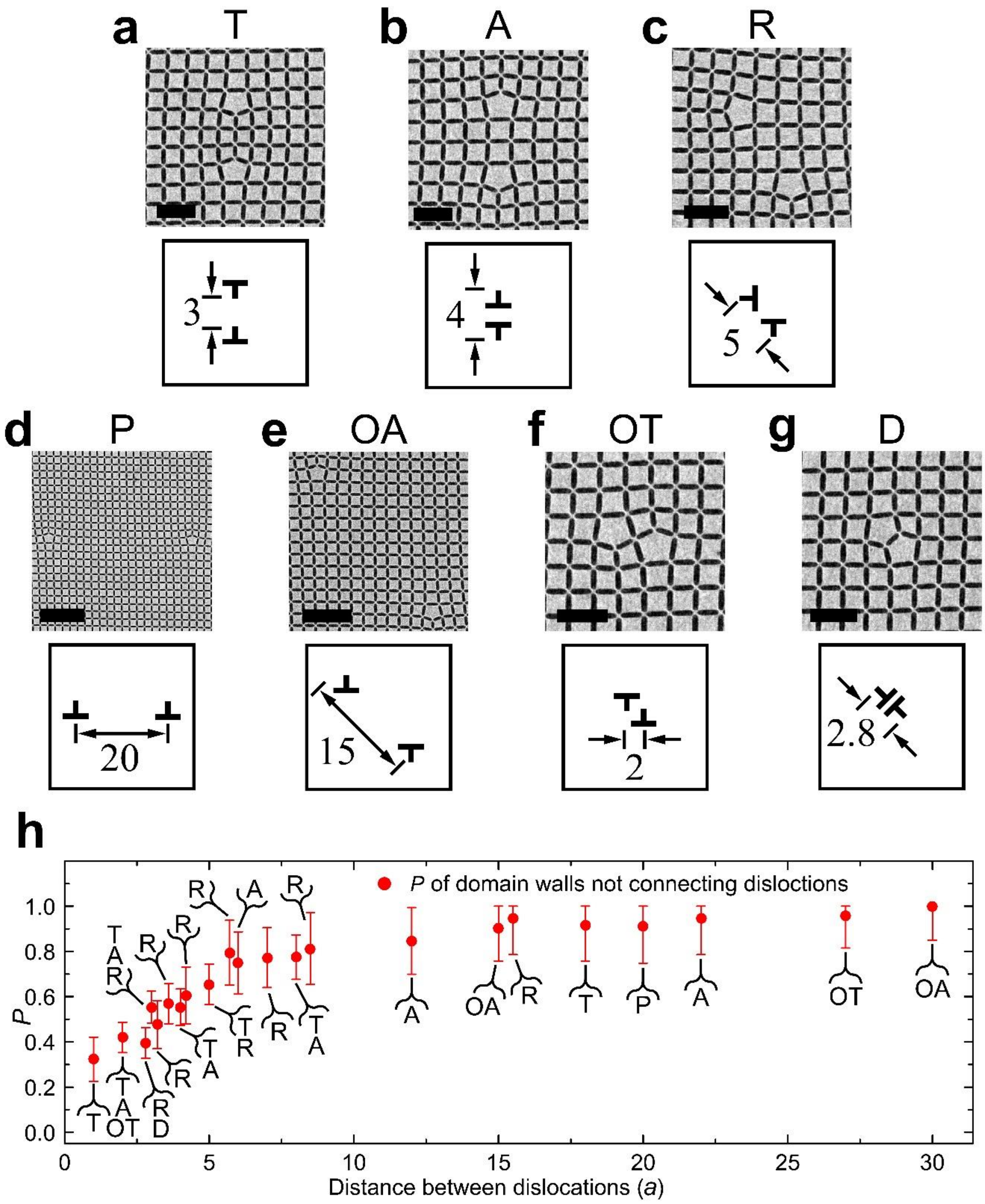

**Supplementary Figure 3 | Geometries studied.** **(a-g)** Definitions for families of two-dislocation geometries studied in this work. Each frame describes one family of geometries where the spacing between dislocations is varied. The spacing is given in units of the lattice constant ($a \approx 500$ nm).

Each frame includes an in-focus TEM image of a representative crystal for that geometry and a schematic representation of the geometry. (**a**) Towards or T geometries have the two dislocations pointing at each other. Scale bar is 1 µm. (**b**) Away or A geometries have the two dislocations pointing away from each other. Scale bar is 1 µm. (**c)** Right or R geometries have the dislocations pointing at right angles to each other. T, A, and R are the most commonly studied geometries. Scale bar is 1 µm. (**d**) Parallel or P geometry has the dislocations pointing in the same direction. Scale bar is 3 µm. (**e**) Opposite-away or OA geometries have the dislocations pointing away from each other and offset. Scale bar is 2 µm. **(f**) Opposite-towards or OT geometries have the dislocations pointing towards each other and offset. Scale bar is 1 µm. (**g)** Diagonal or D geometry has the dislocations pointing diagonally with respect to the crystal axes. Scale bar is 1 µm. (**h**) Probability of domain walls not connecting dislocations vs. distance between dislocations in crystals with two topological defects. Same data as in Fig. 4 in the main text, but with annotations added indicating which geometries are included in each data point. Distances are expressed in units of the lattice constant $a \approx 500$ nm. Error bars are one standard deviation and are calculated from counting statistics. Each geometry includes approximately 40 crystals.